\documentclass[runningheads]{svmult}

\usepackage{makeidx}   % allows index generation
\usepackage{graphicx}  % standard LaTeX graphics tool
\usepackage{subeqnar}  % subnumbers individual equations
\usepackage{multicol}  % used for the two-column index
\usepackage{cropmark} % cropmarks for pages without
\usepackage{physprbb}  % modified textarea for proceedings,
\makeindex             % used for the subject index
\newcommand{\SSS}{\scriptscriptstyle}
\newcommand{\Rl}{R_{\SSS L1}}
\newcommand{\Md}{\dot{\cal M}}
\newcommand{\SC}{\scriptsize}
\newcommand{\Teff}{T_{\mbox{\SC eff}}}
\newcommand{\Tlim}{T_{\mbox{\SC m}}}
\newcommand{\Vkep}{V_{\mbox{\SC kep}}}

\newcommand{\cri}{{\mbox{\tiny crit}}}

\newcommand{\lsim}{{\textstyle{\; \lower 0.7ex\hbox{$<$}\;
  \atop \raise-0.1ex\hbox{$\sim$}}}}
\newcommand{\gcm}{g cm$^{-2}$}

\newcommand{\gs}{gs$^{-1}$}

\begin{document}
\title*{Physical Parameter Eclipse Mapping}
\toctitle{Physical Parameter Eclipse Mapping}
% allows explicit linebreak for the table of content
%
%
\titlerunning{Physical Parameter Eclipse Mapping}
% allows abbreviation of title, if the full title is too long
% to fit in the running head
%
\author{Sonja Vrielmann}
\authorrunning{Sonja Vrielmann}
% if there are more than two authors,
% please abbreviate author list for running head
%
%
\institute{Dept.\ of Astronomy, University of Cape Town,
	Rondebosch, 7700, South Africa}

\maketitle              % typesets the title of the contribution

\begin{abstract}
The tomographic method {\em Physical Parameter Eclipse Mapping} is a
tool to reconstruct spatial distributions of physical parameters (like
temperatures and surface densities) in accretion discs of cataclysmic
variables. After summarizing the method, we apply it to 
multi-colour eclipse light curves
of various dwarf novae and nova-likes like VZ~Scl, IP~Peg in outburst,
UU~Aqr, V2051 Oph and HT~Cas in order to derive the temperatures
(and surface densities) in the disc, the white dwarf temperature, the
disc size, the effective temperatures and the viscosities. The
results allows us to establish or refine a physical model for the
accretion disc. Our maps of HT~Cas and V~2051Oph, for example, indicate
that the (quiescent) disc must be structured into a cool, optically
thick inner disc sandwiched by hot, optically thin chromospheres. In
addition, the disc of  HT~Cas must be patchy with a covering factor of about
40\% caused by magnetic activity in the disc.
\end{abstract}

%%%%%%%%%%%%%%%%%%%%%%%%%%%%  Introduction  %%%%%%%%%%%%%%%%%%%%%%%%%%%%

\section{Introduction}
Cataclysmic variables (CVs) are close interacting binary system
consisting of a main sequence star (the secondary) and a white
dwarf. The Roche-lobe filling secondary loses matter via the inner
Lagrangian point to the primary. If the magnetic field of the white
dwarf is negligible, then angular momentum conservation drives the
matter from the secondary into an accretion disc around the primary
component.  This disc matter is maintained by a steady mass stream
from the secondary, which hits the disc in the so-called bright spot.

Some of these systems undergo frequently an outburst, in
which the system brightens for a short period of time (a few days) in
comparison to the interval between such eruptions (few weeks to
months). It is believed that the accretion disc follows a disc
instability cycle \cite{O} in which the hydrogen in the disc
switches its ionization status. Other systems, the so-called
nova-like variables, appear to be in a permanent outburst state and are believed
to exhibit optically thick, steady state discs. Quiescent disc, in
contrast, show line emission, mainly of hydrogen and are therefore at
least partially optically thin. An extensive overview over the
theoretical and observational aspects of CVs is given in Warner
\cite{W95}.

The ultimate wish of any astronomers is probably to be able to see the
stars from close by, to get a direct image of the stellar
systems. Tomographic methods give us a unique opportunity to
get such a picture, even if indirect, without getting close to the
object. By making use of the eclipse light curve, the classical
Eclipse Mapping technique (EM, \cite{H85}) provides us with images of the
accretion disc, while Physical Parameter Eclipse Mapping (PPEM,~\cite{VHH})
give us some insight into the accretion
physics. This paper gives an introduction to PPEM and summaries
applications of this method to multi-colour light curves of five
different cataclysmic variables.

%%%%%%%%%%%%%%%%%%%%%%%%%%%%%%%  PPEM  %%%%%%%%%%%%%%%%%%%%%%%%%%%%%%%%

\section{The PPEM method}
\label{ppem}

\begin{figure}
\begin{center}
\includegraphics[width=.7\textwidth]{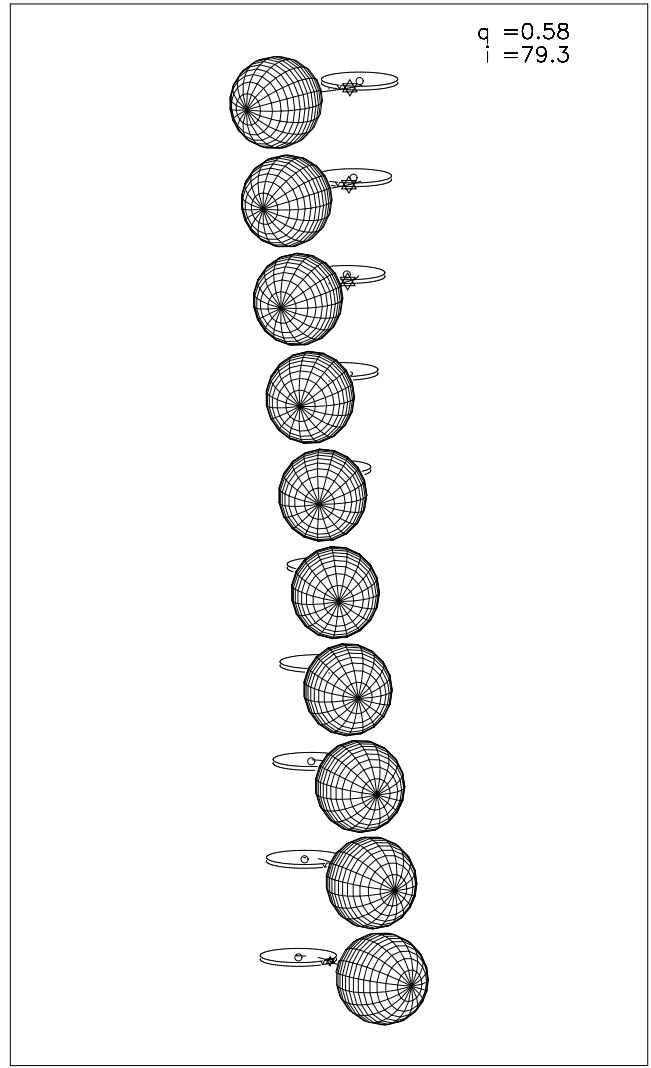}
\end{center}
\caption[]{Illustration of a cataclysmic variable going through an
eclipse of the accretion disc. The parameters are set for IP~Peg.
Courtesy to K.~Horne for the program {\em cvmovie}
\label{fig_cvmovie}}
\end{figure}

Figure~\ref{fig_cvmovie} illustrates the varying viewing angle a high
inclination CV undergoes during an eclipse. One can see that at any
one time during eclipse, only part of the accretion disc is occulted
by the secondary.  The flux at the given orbital phase is the total of
all emission of any material not eclipsed at this phase. Any local
intensity maximum in the accretion disc, e.g.\ the bright spot, will
be eclipsed and reappears at a characteristic orbital phase and causes
thereby a pair of steep gradients or steps in the observed eclipse
light curve, one in ingress and the other in egress. Depending on the
spatial location of this spot, the pair of steps will be shifted in
relation to phase 0 which is defined as the conjunction of the white
dwarf.

In EM one takes advantage of this spatial information of
the intensity distribution of the accretion disc hidden in the eclipse
profile. We have to make three basic assumptions: (a) the geometry of the
secondary is known (it is usually a good approximation to assume it
fills its Roche lobe); (b) the geometry of the disc is known (in the
simplest approximation we use a geometrically infinitisimally thin
disc); (c) the disc emission does not vary with time (this can be
achieved by using averaged eclipse profiles in order to reduce the
amount of flickering and flares from the disc).
By fitting the eclipse light curve, we reconstruct the intensity
distribution using a maximum entropy method (MEM, \cite{SB}).
The MEM algorithm allows one to choose the simplest solution
still compatible with the data in this otherwise ill-conditioned back
projection problem. Further details of the EM method can be found in
Horne \cite{H85} or Baptista \& Steiner \cite{BS}. The usefulness of this
method is extensively described by R. Baptista in these proceedings.

\begin{figure}
\begin{center}
\includegraphics[width=.9\textwidth]{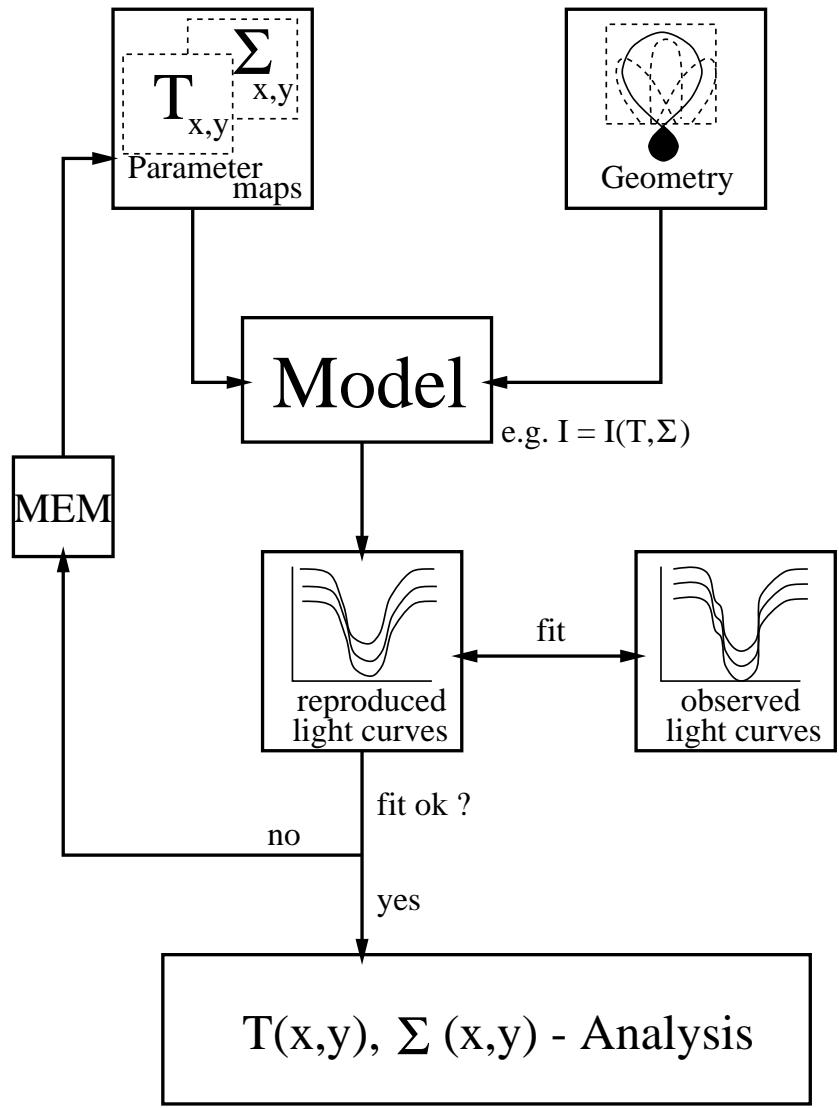}
\end{center}
\caption[]{A flowchart diagram to illustrate the Physical Parameter
Eclipse Mapping algorithm
\label{fig_ppem}}
\end{figure}

The PPEM approach goes a step further in that we map physical
parameters, like temperature $T$ and surface density $\Sigma$, instead
of intensities. Figure~\ref{fig_ppem} explains the method by means of
a flowchart diagram. In addition to the three above mentioned basic
assumptions (indicated by the right upper panel) we have to presuppose
a spectral model for the disc emission (indicated as ``Model'' in the
chart), relating the parameters to be mapped (e.g.\ $T$, $\Sigma$) to
the radiated intensity in a given filter (e.g.\ $I_\nu = f(T,
\Sigma)$). The model can be as simple as a pure black body spectrum,
the PPEM is then called {\em Temperature Mapping} or as complicated as
non-linear LTE disc model atmospheres as calculated by Hubeny
\cite{H91}. Two simple model spectra are described in
Sect.~\ref{model}. The predicted multi-colour light curves are then
compared to the observed ones. As long as the fit is not satisfactory,
the parameter maps are varied according to the MEM algorithm (i.e.\
using gradients $\delta I/\delta T$ (and $\delta I/\delta
\Sigma$)). As soon as the $\chi^2$ that was aimed for is reached (and
the maps have converged, see \cite{H85}) the maps $T$ (and $\Sigma$)
can be further analysed and compared to current disc models (see
Sect.~\ref{expect}).

%%%%%%%%%%%%%%%%%%%%%%%%%%  Light curves  %%%%%%%%%%%%%%%%%%%%%%%%%%%

\subsection{The light curves}

The PPEM method requires input in the form of light curves. The number
of light curves in various wavelength regions (filters/passbands)
necessary to calculate a reliable parameter map depends on the used
spectral model (see following Sect.~\ref{model}) and must be equal
to or greater than the number of parameters to be mapped. For example,
if one uses a black body spectrum to map the temperature (see
Sect.~\ref{ippeg} or \ref{vzscl}), one light curve would be
sufficient for a unique PPEM analysis. However, the more light curves
(in various filters) one uses the better: the long wavelength data
(e.g.\ IR light curves) will be ideal to determine low temperature
(outer) regimes in the disc, while the short wavelength data (e.g.\ UV
light curves) will be most appropriate to map the hot (inner) disc
regions.

If one uses an appropriate spectral model, such as calculated by
Hubeny~\cite{H91}, spectrophotometry can be analysed. PPEM ideally
allows to extract a maximum of the information content of the data.

%%%%%%%%%%%%%%%%%%%%%%%%%%%%%%  Model  %%%%%%%%%%%%%%%%%%%%%%%%%%%%%%%

\subsection{The model spectra}
\label{model}
The spectral model provides a relation between the mapped parameter(s)
$P$ and the intensity $I_\nu$ at the frequency $\nu$. If one would set
$P = I_\nu$, this simply reduces PPEM to EM. In the simplest true PPEM
version we use a black body spectrum for the spectral model, i.e.
\begin{equation}
I_\nu = B_\nu(T) = \frac{2 h \nu^3}{c^2} \left(e^{\frac{h \nu}{kT}}-1\right)
%	B_\nu(T) = \frac{2 h \nu^3}{c^2} \frac{n_\nu}{e^{\frac{h \nu}{kT}}-1}
\end{equation}
which can be used to map the parameter temperature $T$ in
the disc. Here $c$ is the speed of light, $h$ Planck and $k$ Boltzmann
constant. Using this model we assume that the disc radiates optically
thick. This option of the PPEM method is called {\em Temperature
Mapping} and is useful for accretion discs in novae, nova-like variables and
outbursting dwarf novae.

In order to allow the disc to be optically thin, as we expect in
quiescent dwarf novae, we introduce a second parameter. We have some
freedom in the choice, but since theoretical studies e.g.\ of the
outburst behaviour use the surface density $\Sigma$ in the disc, we
use the same. The next simple model thus is calculated as:
\begin{equation}
I_\nu = B_\nu(T) (1 - e^{-\tau/\cos i})
\end{equation}
with the optical depth
\begin{equation}
\tau = \rho\, \kappa\, 2 H,
\end{equation}
where $\kappa$ is the mass absorption coefficient,
$\rho$ the mass density
\begin{equation}
\rho = \Sigma\, /\, 2 H
\end{equation}
and $H$ the disc scale height
\begin{equation}
H/R = c_s/\Vkep.
\label{eq_hr}
\end{equation}
Here, $R$ is the radius, $c_s = \sqrt{kT/(\mu m_H)}$ is the local sound
speed and $\Vkep = \sqrt{G {\cal M}_1/R}$ the Keplerian velocity of
the disc material, with $\mu$ the mean molecular weight, $m_H$ the
mass of a hydrogen atom, $G$ the gravitational constant and ${\cal M}_1$
the mass of the white dwarf.

In our studies we used a pure hydrogen slab in local thermodynamic
equilibrium (LTE) including only bound-free and free-free H and H$^-$
emission. Even though this model is very simple it still is very
useful in that it allows us to distinguish between optically thin and
thick regions of the disc. The eclipse mapping of dwarf nova discs and
the presence of emision lines suggest that at least part of the
quiescent discs are optically thin.

For the usual range of temperatures and surface densities in CV
accretion discs, this LTE approximation is still relatively good. In
more realistic models (e.g.\ models by \cite{H91}) one would
include metals which significantly increase the opacity at
temperatures lower than about $\Tlim \sim 6300$K leading to lower
intensities than determined by our $I(T,\Sigma)$. Therefore, using our
simple model, our algorithm overestimates the temperatures (and if
optically thin also the surface density) in the cooler parts of the
disc, in order to meet the required intensities. This will only be
important in the outer regions of the disc with temperatures below
$\Tlim$.

%%%%%%%%%%%%%%%%%%%%%%%%%%%%  White Dwarf  %%%%%%%%%%%%%%%%%%%%%%%%%%%%

\subsection{The white dwarf}
\label{sec_wd}
For the white dwarf emission we used white dwarf spectra for the
temperature range 10\,000~K to 30\,000~K. Outside this range black
body spectra are used. The white dwarf is assumed as a spherical
object in terms of occultation of the accretion disc and eclipse by
the secondary, but it is attributed only a single temperature.  Since
the absorption lines partly coincide with the mean wavelengths of the
filters, pass band response functions are used. Usually, the lower
hemisphere of the white dwarf is assumed to be obscured by the inner
disc.

%%%%%%%%%%%%%%%%%%%%%%%%%%%%%%%  Grid  %%%%%%%%%%%%%%%%%%%%%%%%%%%%%%%

\subsection{The spatial grid for the accretion disc}
In the present studies the accretion disc is assumed to be
infinitesimally thin. Onto this disc we constructed a two-dimensional
grid of pixel. This grid consists of rings cut into a number of pixel
that provides equal areas for all pixel \cite{VHH}.
For reference of spatial structures in the disc, we use radius
and azimuth. The radius is used in units of the distance between the
white dwarf and the inner Lagrangian point $L_1$, 
the white dwarf being at the origin.
%i.e.\ the point $L_1$ has a radius of $1\Rl$.  
The second coordinate is the azimuth,
the angle as seen from the white dwarf and counting from $-180^\circ$
to $180^\circ$. Azimuth 0 points towards the secondary, the leading
lune of the disc has positive and the following lune negative azimuth
angles.

%%%%%%%%%%%%%%%%%%%%%%%%%%%%  uneclipsed  %%%%%%%%%%%%%%%%%%%%%%%%%%%%

\subsection{The uneclipsed component}
Apart from the disc emission we allow for another emission component
that is never eclipsed, e.g.\ the secondary. However, we do not need
to specify a geometrical location or a spectral model for this
uneclipsed component, but can reconstruct it for each wavelength/pass
band independently as a constant flux contribution. It is usually
given in the light curve plots as a solid horizontal line.

The uneclipsed component can then be used to give limits on the
secondary or, if its contribution is known, the remaining uneclipsed
flux can be analysed. It can originate in disc regions never eclipsed
due to a relatively low inclination angle, or more likely regions at
heights $z$ above the disc that are never eclipsed by the secondary.

%%%%%%%%%%%%%%%%%%%%%%%%%%%%%%  Distance  %%%%%%%%%%%%%%%%%%%%%%%%%%%%%%

\subsection{The distance to the systems}
\label{distance}
In order to reconstruct sensible physical parameter distributions within the
accretion disc, the distance to the system has to be known. Only in some
cases, relatively good estimates could be made using e.g.\ the secondary
absorption line spectrum or the white dwarf flux. 

PPEM provides also the option to determine an independent estimate of
the distance, using the combined flux of the accretion disc and the
white dwarf. This estimate can either be compared to present estimates
(e.g.\ HT Cas, UU~Aqr) or give the first opportunity to establish a
distance (e.g.\ V2051~Oph).

The PPEM distance estimate is based on the fact that a spectrum with
at least 3 wavelength points is not only determined by the 
temperature and the surface density, but also by the distance. If the
assumed distance is smaller or larger than the true distance, the fit
to the data will be poor. In a multi-pixel analysis like PPEM this
might be compensated by neighbouring pixel leading to an overall
``good'' fit to the observed light curve (expressed as a small
$\chi^2$), but the resulting reconstruction will show artificial
structures and the predicted light curve will have kinks not justified
by the data. The amount of structure is expressed as the entropy $S$
of the map: the smoother the map the higher the entropy. For each
trial distance $d$ we converge the maps to a specific $\chi^2$ and
then plot the parameter $S$ against $d$. Where this function peaks,
the map shows the least amount of structures and we will ideally find
the true distance.

Only maps corresponding to fits of equal goodness can be compared for
this distance estimate, since the entropy not only depends on the
distance but also on the final $\chi^2$. The lower the $\chi^2$, i.e.\
the better the fit, the more structures will appear in the map,
reducing thereby the entropy. At which $\chi^2$ to stop has to be
determined individually, it depends on the goodness of the spectral
model or the amount of flickering in the light curve and is therefore a
somewhat subjective measure.

Using this distance estimate one presupposes that the spectral model
describes the true emissivity of the disc reasonably well. It is
difficult to determine the error introduced by a wrong model. A
generally good fit to the observed multi-colour eclipse light curves,
however, seems to justify the model used.

%%%%%%%%%%%%%%%%%%%%%%%%%%%%%%%  Tests  %%%%%%%%%%%%%%%%%%%%%%%%%%%%%%%

\section{The reliability of the PPEM method}
\label{tests}

\begin{figure}
\begin{center}
\includegraphics[width=.8\textwidth]{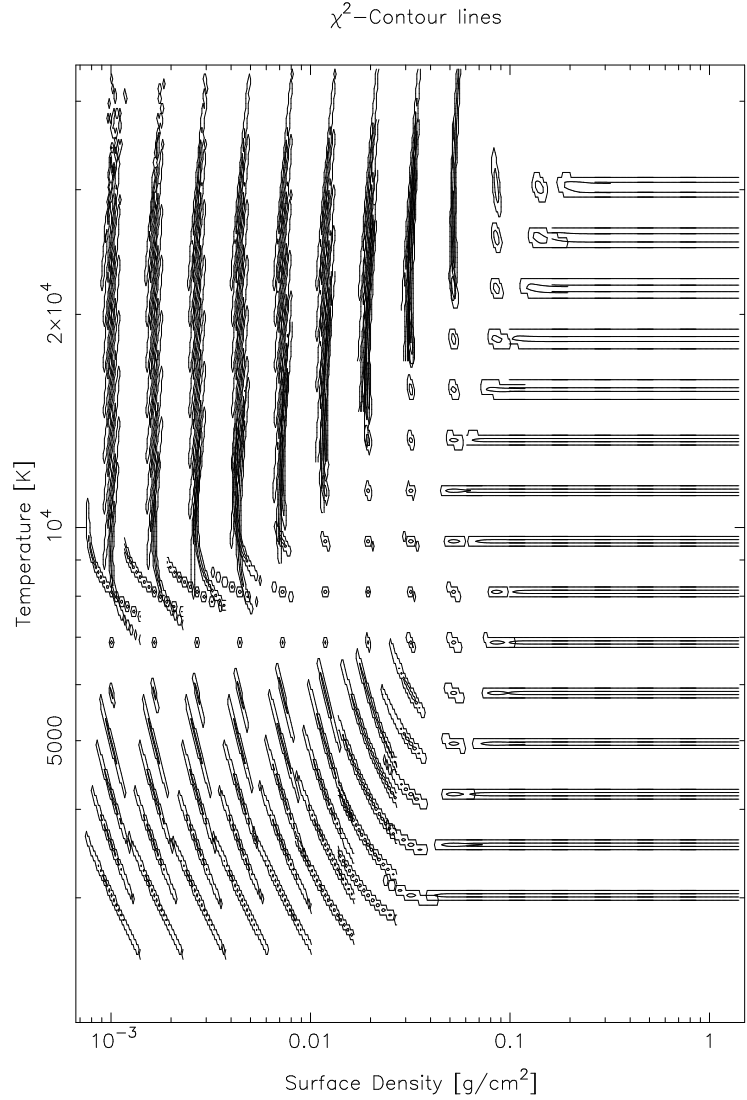}
\end{center}
\caption[]{$\chi^2$ contours for different parameter combinations of
temperature $T$ and surface density $\Sigma$ for a disc radius $R =
0.32\Rl = 1 \cdot 10^{10}$ cm. For each parameter pair $T,\Sigma$ the
spectrum $I(T,\Sigma)$ was calculated, then the spectra for
$I(T\pm\Delta T,\Sigma\pm\Delta \Sigma)$ with small values for $\Delta
T$ and $\Delta \Sigma$ and the difference in the spectra expressed as
the $\chi^2$. Contours are drawn for $\chi^2 = 3$ and 1 assuming an
error in the spectra of 1\% of the maximum value. The narrower the
contour line, the better determined the parameter.
\label{fig_xi2}}
\end{figure}

Tests of the PPEM method were shown in Vrielmann et al.\cite{VHH}. In
general, the method allows to reconstruct the parameter distributions
very well. However, it depends on the spectral model and the parameter
values, how reliable the resulting maps are.

The black body spectrum is a non-linear function of the temperature,
but since it gives an unambiguous function $I(T)$ it can be used to map
the temperature uniquely. Therefore, the {\em Temperature Mapping}
gives very reliable results. Since PPEM is using multi-colour light
curves we see a huge improvement compared to classical eclipse
mapping: steep gradients (e.g.\ at the disc edge) are much better
reproduced in PPEM \cite{VHH}.

For a $T-\Sigma$ model as described in Sect.~\ref{model} one has to
check where in the parameter space the solutions $I(T,\Sigma)$ are
unique before the maps can be reliably analysed. For example, in the
optically thick limit (large $\Sigma$), the surface density can assume
any value above a certain limit without a change in the spectrum.
However, the temperature in this case is very well defined. In other
regions of the parameter space only the surface density is well
defined (large $T$, small $\Sigma$), in again other regions both
(intermediate values of $T$ and $\Sigma$) or none (see the {\em
banana} shaped contour lines for small $T$, small $\Sigma$ in
Fig.\ref{fig_xi2}) of the parameters can be determined uniquely. This
pattern changes slightly with the disc radius. Figure~\ref{fig_xi2}
shows this pattern for a radius of $R = 0.32\Rl = 1 \cdot 10^{10}$ cm.

Parameters are most reliable in disc regions that emit the
largest intensities. In disc regions with very low intensities the
reconstructed parameters may be influenced by the MEM algorithm
leading to a smoothing of the spatial gradients. This is typically the
case in the outer disc regions.

Such a study can be used to derive error bars for the reconstructed
temperature and surface density distribution. It is usually sufficient
to determine these errors for azimuthally averaged maps as an
indication for the reliability of the parameter values.

A useful tool to analyse the derived maps is the ratio of the averaged
intensity distribution $I(T,\Sigma)$ to the black body intensity
distribution $I_{BB}(T)$ derived from the temperature alone. In the
optically thick case and for very low intensities this ratio reaches
unity.

%%%%%%%%%%%%%%%%%%%%%%%%%%%%%%%  Expect  %%%%%%%%%%%%%%%%%%%%%%%%%%%%%%%

\section{What to expect?}
\label{expect}
The reconstructed parameter distributions will allow us to calculate
further parameters, like the optical depth, the disc scale height, the
effective temperature and mass accretion rate or the viscosity. These
gives us clues about the physics going on in the disc. What we can
expect is described for the reconstructed and a few derived
parameters.

\subsection{The gas temperature}

The gas temperature is expected to rise towards the disc centre. Close
to the white dwarf the disc material is accelerated and the
gravitational energy release is strongest. If the accretion disc emits
as a black body, the gas temperature is equal to the effective
temperature (see Sect.~\ref{sec_teff}) and should decrease radially
roughly like $T \sim r^{-3/4}$. In the optically thin case, the
temperature distribution may be much flatter than this.

In case the disc has a hole, the reconstructed temperature values will
decrease towards the disc centre below a value that is undetectable
at the wavelengths used. However, if the hole is very small, MEM will
lead to a smearing of the values resulting in a flat temperature
distribution at small radii.

\subsection{The surface density}
The literature is divided about the radial behaviour of the surface
density. While Meyer \& Meyer-Hofmeister's \cite{MM} calculations result
in a radially decreasing surface density distribution, Ludwig,
Meyer-Hofmeister \& Ritter \cite{LMR} and Cannizzo et al.~\cite{CGW}
derive slightly increasing surface density distributions.

Since the critical surface density distributions within the disc
instability model show an increase of $\Sigma_\cri$ with radius
\cite{CW,HLD} we would rather expect that the actual $\Sigma(r)$
distribution also increases with radius. This would make it much
easier for a disc to undergo an outburst.

Note, that if the disc is optically thick, we have no means to
determine the surface density, only a lower limit that may be far from
the true value.

\subsection{The effective temperature}
\label{sec_teff}
Considering the gravitational energy release and taking into account the
slowing down of the disc material at the white dwarf surface, a steady
state accretion disc should have the following radial effective
temperature distribution \cite{FK}:

\begin{equation}
\label{eq_teff}
\Teff(r) = T_* \left(\frac{r}{R_1}\right)^{-3/4} \left[1 -
	\left(\frac{R_1}{r}\right)^{1/2}\right]^{1/4}
\end{equation}
with
\begin{equation}
T_* = \left[\frac{3G{\cal M}_1\Md}{8\pi \sigma R_1^3}\right]^{1/4}
\end{equation}
where $r$ is the radius in the disc, $R_1$ and ${\cal M}_1$ are the
radius and mass of the white dwarf, $G$ the Gravitational constant,
$\Md$ the mass accretion rate of the disc and $\sigma$ the
Stephan-Boltzmann constant. For large radii Eq.~(\ref{eq_teff}) reduces to
\begin{equation}
\Teff(r) = T_* \left(\frac{r}{R_1}\right)^{-3/4} \,\,\,\, \mbox{for}
\,\,\,\, r\gg R_1.
\end{equation}

Deviations from this distribution in real disc show that it is not in
steady state. This can be expected for a quiescent accretion disc of a
dwarf nova awaiting the next outburst. Discs in nova-like variables
and in outbursting dwarf novae are expected to follow the steady state
$\Teff(r)$-profile.

\subsection{The critical temperature}
The effective temperature distribution can then also be compared to a
critical temperature, as e.g.\ calculated by Ludwig et al.~\cite{LMR}. If
the effective temperature distribution falls below this critical value
the accretion disc is in the lower branch of the $\Teff-\Sigma$
hysteresis curve which is used to explain the disc instability
cycle. Such a disc should undergo outbursts. If the temperatures are
above this critical value, the accretion disc is on the hot branch of
this hysteresis curve and should therefore be currently in outburst or
within a nova-like or nova system.

\subsection{The viscosity}
\label{viscosity}
A further parameter that can be derived is the viscosity $\nu$,
usually parametrized as $\nu = \alpha c_s H$ (where $c_s$ is the
local sound speed and $H$ again the disc scale height) using Shakura
\& Sunyaev's \cite{SS} $\alpha$-parameter. Theoretical studies involving
the disc instability cycle predict values of $\alpha$ around 0.01 in
quiescence \cite{S84}.

We calculate the viscosity using the standard relation between the
viscously dissipated and total radiated flux $F_\nu$,
\begin{equation}
\label{eq_viscosity}
2 H \alpha P \frac{3}{4} \Omega_K = \int F_\nu d \nu = \sigma \Teff^4
\end{equation}
where $P$ is the (gas) pressure and $\Omega_K$the Keplerian angular
velocity of the disc material.

\section{Application to real data}
This section gives some highlights of the application of the PPEM
method to different objects. Any details about the analyses can be
found in the corresponding, mentioned articles.

%%%%%%%%%%%%%%%%%%%%%%%%%%%%%%%  VZ Scl  %%%%%%%%%%%%%%%%%%%%%%%%%%%%%%%
\subsection{VZ~Sculptoris}
\label{vzscl}

VZ~Scl is a little studied eclipsing nova-like with an extreme
difference between high (normal) and low states of about 4.5 mag
\cite{OFW}. Since in the low state the secondary dominates the
spectrum at all wavelengths, Sherington et al.~\cite{SBJ} could
determine a reliable distance of 530~pc. The other system parameters,
in particular the inclination angle $i$ and the mass ratio $q$ are
somewhat uncertain. We adopted the same values O'Donoghue et
al.~\cite{OFW} used for their Eclipse Mapping and the ephemeris
determined by Warner \& Thackeray~\cite{WT}.

\begin{figure}
\begin{center}
\includegraphics[width=.5\textwidth]{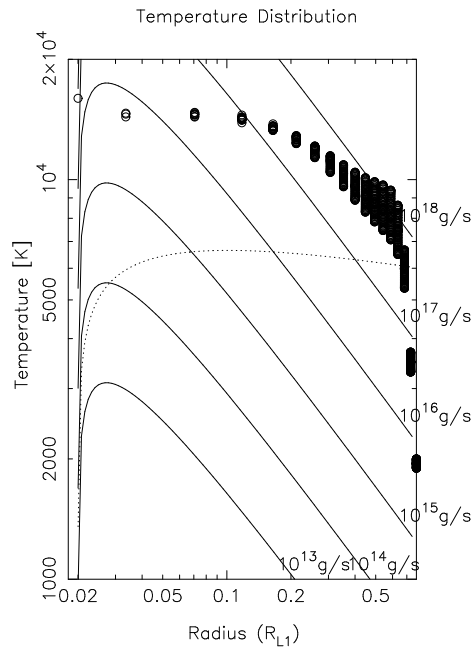}
\end{center}
\caption[]{The temperature distribution of {\bf VZ~Scl}.
\label{vz_t}}
\end{figure}

Our observations were according to O'Donoghue et al.~\cite{OFW} taken
when the object was in its normal state and we therefore used the {\em
Temperature Mapping} version of PPEM \cite{V99}. As expected, the
black body assumption turned out to be a fairly good approximation. We
reached a $\chi^2$ of 5. The disc appears to be in steady state
between $0.15~\Rl$ to the disc edge at $0.7~\Rl$ with a mass accretion
rate of about $3\times10^{17}$\gs (see Fig.~\ref{vz_t}).  In the inner
part of the disc ($r < 0.15~\Rl$) the temperature profile is very
flat. This could be caused by a small hole in the disc, that is
smeared out due to the MEM algorithm. Other explanations are discussed
by e.g.\ Rutten et al.\ \cite{RvT}, but the true cause is still not
clear.

Note in particular, that we simply used the published distance and
still achieved a relatively good fit to the data.  In a future more
extensive study we will test if the disc has optically thin region and
if we can reach a better fit to the data using the $T-\Sigma$ version
of PPEM. We will then also determine an independent distance
estimate. However, we do not expect a great deviation from the
literature value.

%%%%%%%%%%%%%%%%%%%%%%%%%%%%%%%  IP Peg  %%%%%%%%%%%%%%%%%%%%%%%%%%%%%%%
\subsection{IP~Pegasi}
\label{ippeg}

\begin{figure}
\begin{center}
\includegraphics[width=.9\textwidth]{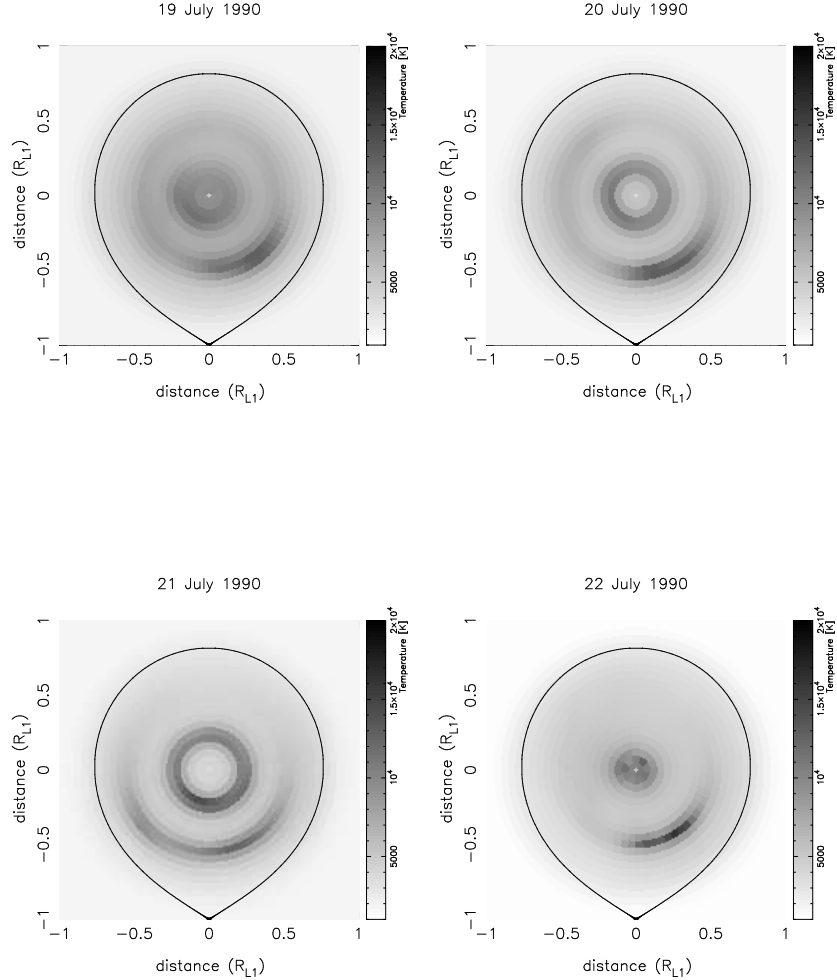}
\end{center}
\caption[]{The temperature maps of {\bf IP~Peg} on decline from
outburst. The secondary is always at the bottom, just outside the
plotted area.
\label{ip_m}}
\end{figure}

IP~Peg is probably the best studied dwarf nova, and still is not
completely demystified. It is most famous for the spiral waves found
by Steeghs, Harlaftis, Horne (\cite{SHH}, see also these proceedings)
during rise to outburst.

We analysed UBVRI data taken during four nights on decline from
outburst \cite{BHMW} by feeding them into our PPEM
algorithm \cite{V97} using the {\em Temperature Mapping}
option. We achieved relatively good fits to the data, with $\chi^2$'s
ranging between 1.1 and 4.

All four temperature maps (Fig.~\ref{ip_m}) show a prominent
azimuthally smeared out bright spot at a radius of 0.5 to
0.55$\Rl$. The temperature in the central part, up to a radius of
0.1$\Rl$ drops dramatically during the three nights after maximum
light from about 9\,000~K to 4\,000~K, possibly indicating an emptying
out of the inner disc. Only in the last of the four nights, when the
system reached the quiescent brightness, does the temperature reach
again about 10\,000~K at small radii. Instead, a hot ring at radius
0.2$\Rl$ present in the first three nights has disappeared in the last
one. This mysterious behaviour seems to indicate that the outburst
takes place only in the inner parts of the disc, without evidence of a
cooling front.

These studies of VZ~Scl and IP~Peg during decline from outburst show
that even the simple {\em Temperature Mapping} is a very useful tool.

\begin{figure}
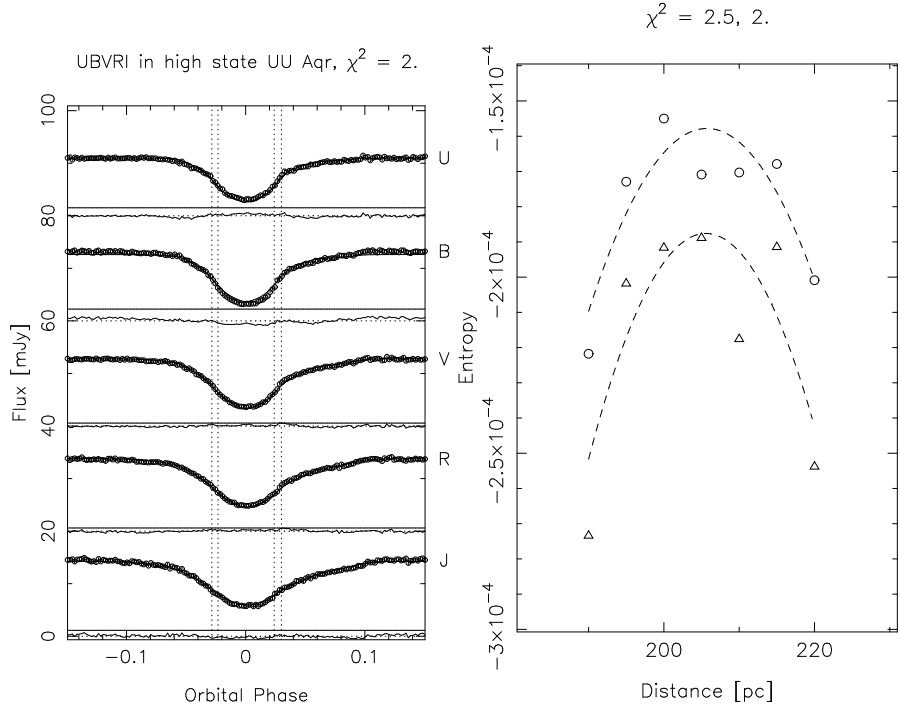

\begin{center}
\includegraphics[width=.48\textwidth]{uuh205_2f.epsi}
\includegraphics[width=.48\textwidth]{uuh206_2en.epsi}
\end{center}
\caption[]{{\em Left:} Averaged UBVRI light curves of {\bf UU~Aqr} in
high state with the PPEM fits. The light curves are shifted upwards
with the dotted line giving the zero line for each light curve. The
residuals and the uneclipsed component are plotted in relation to the
zero lines. Vertical dashed lines give the phases of the white dwarf
ingress and egress. {\em Right:} Entropy vs.\ trial distance for
$\chi^2 = 2.5$ (circles) and $\chi^2 = 2$ (triangles) and parabolic
fits to the points peaking at 206~pc.
\label{uu_f}}
\end{figure}

\begin{figure}
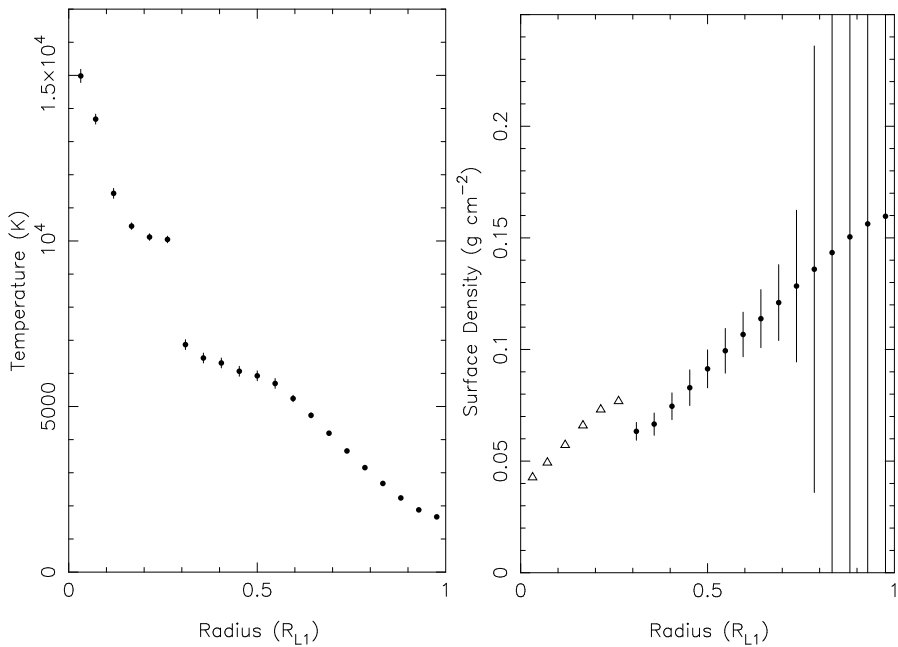

\begin{center}
\includegraphics[width=.48\textwidth]{uuh205_2a1.epsi}
\includegraphics[width=.48\textwidth]{uuh205_2a2_2.epsi}
\end{center}
\caption[]{{\bf UU~Aqr:} Azimuthally averages of the reconstructed
temperature ({\em left}) and surface density ({\em right})
distributions. The error bars give an indication of the reliability of
the reconstructed values according to a study of the spectral model.
In the inner part is the disc optically thick, therefore the surface
density there gives only lower limits (indicated by triangles). In the
outer regions, the intensity $I(T,\Sigma)$ becomes very small, leading
to large error bars in $\Sigma$.
\label{uu_p}}
\end{figure}

%%%%%%%%%%%%%%%%%%%%%%%%%%%%%%%  UU Aqr  %%%%%%%%%%%%%%%%%%%%%%%%%%%%%%%
\subsection{UU~Aqarii}
\label{uuaqr}
The eclipsing nova-like UU~Aqr belongs to the SW~Sex stars, i.e.\ in
spite of the high inclination it shows single peaked emission lines
with phase dependent absorption features. Doppler tomography reveals a
disc, with the asymmetric part of the emission lines caused by a
prominent bright spot \cite{HSSSB}. This disc shows
long term photometric variations in the form of high and low states
with an amplitude of about 0.3~mag.

Baptista et al.~\cite{BSC} repeatedly observed UU~Aqr over a period of
about 6 years catching the system in a high and two low states. We
applied the PPEM method to their averaged light curves separated in a
high state and a low state light curve. Here, we present only the
application of PPEM to the averaged high state data. An analysis of
the averaged low state data (and of the individual light curves) will be
presented by Vrielmann~\cite{V20U}.

Figure~\ref{uu_f} (left) shows the averaged high state UBVRI light curves
together with the PPEM fits. Baptista et al.~\cite{BSC} estimated a
distance to UU~Aqr by fitting the white dwarf fluxes. If the inner
disc is opaque and obscures the lower hemisphere of the white dwarf,
they derive a distance of $270\pm50$~pc. Assuming an optically thick
inner disc, Baptista et al.~\cite{BSH} performed a cluster main sequence
fitting similar procedure to derive a distance of $200\pm30$~pc.

\begin{figure}
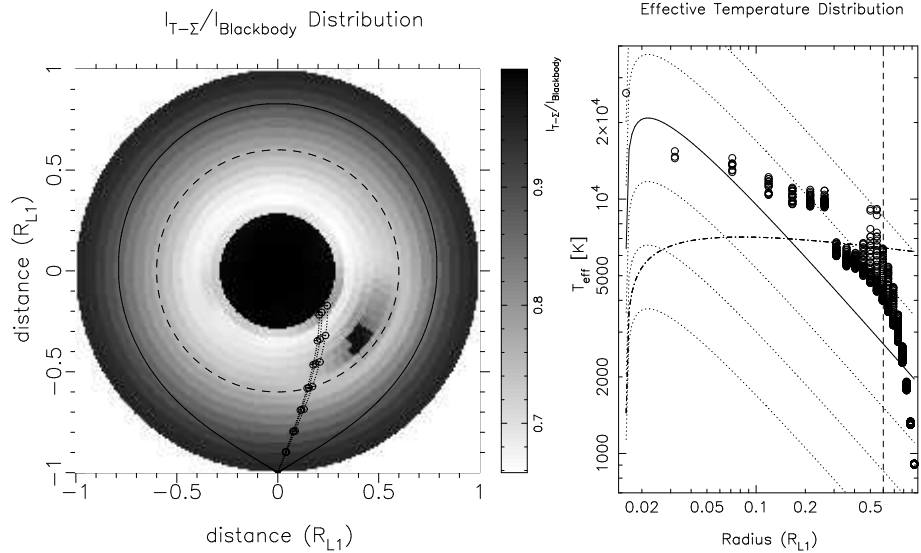

\begin{center}
\includegraphics[width=.6\textwidth]{uuh205_2iv.epsi}
\includegraphics[width=.38\textwidth]{uuh205_2t.epsi}
\end{center}
\caption[]{{\bf UU~Aqr:} {\em Left:} The azimuthally and spectrally
averaged intensity ratio $I(T,\Sigma)/I_{BB}(T)$ as explained in
Sect.~\ref{tests} as a gray-scale plot. The dashed circle is the
disc radius, the pear shaped line is the Roche-lobe of the primary and
the lines from the inner Lagrangian point are ballistic stream lines
for mass ratios $q\pm50\%$. The secondary is at the bottom. {\em
Right:} The radial effective temperature distribution. Underlying
dotted lines are effective temperature distributions for steady state
discs with mass accretion rates $\log \Md = 13$ to 18, the one for
$\Md = 10^{16}$gs$^{-1}$ is drawn solid for reference. The dashed
vertical line gives the disc radius, the dash-dotted line a critical
temperature according to Ludwig et al.~\cite{LMR}
\label{uu_teff}}
\end{figure}

We independently estimated the distance to UU~Aqr using the PPEM
method as described in Sect.~\ref{distance}. Figure~\ref{uu_f} (right)
shows the entropy as a function of the trial distance. The parabolic
fit to the data for $\chi^2=2$ peaks at a distance of 206~pc,
consistent with Baptista et al.'s \cite{BSH} estimate.

Fig.~\ref{uu_p} shows the averages of the reconstructed temperature
$T$ and surface density distributions $\Sigma$ with error bars
indicating the reliablility of the reconstructed values. The derived
intensity distribution $I(T,\Sigma)$ for the filter~I helps us to
define a disc radius. $I(T,\Sigma)$ has values at $r=0.6\Rl$ of less
then 10\% of the white dwarf value. In the outer regions the intensity
vanishes and therefore the Balmer Jump disappears and the error bars
of $\Sigma$ again become very large.  For large radii, the temperature
drops below $\Tlim$, i.e.\ the true temperatures are even lower and
reach undetectable limits. Finally, the original $T, \Sigma$ maps show
a bright spot at a radius of $r_{\mbox{\footnotesize spot}}=0.6\Rl$
(see also Fig.~\ref{uu_teff}, left) which lets us set the disc size as
$0.6\Rl$.

While the temperatures are everywhere well defined and decrease with
radius as expected, the surface density values are only in the outer
parts of the disc ($0.3\Rl < r < 0.6\Rl$) trustworthy. They clearly
increase with radius even within the allowance of the error bars. In
the inner part the disc is optically thick (see Fig.~\ref{uu_teff},
left) and we can only derive lower limits for $\Sigma$. The true
values, however, may be much larger than those limits.

The white dwarf temperature was reconstructed to 26\,100~K using white
dwarf spectra as described in Sect.~\ref{sec_wd}.  This value very
well lies in the usual range of white dwarf temperature estimates of
between 9\,000~K and 60\,000~K \cite{S98} and quite close to the average
effective temperatures of white dwarfs in non-magnetic CVs, $<\Teff> =
24\,100$~K~ \cite{S99}.

As Fig.~\ref{uu_teff} (left) shows, only the central region up to
about $0.3\Rl$ is optically thick, the remaining disc is optically
thin, including the bright spot. This is unexpected, nova-likes are
expected to have overall optically thick discs. However, this
indicates a probable location for the line emission, i.e.\ in the
outer parts of the disc.

The gray-scale plot shows that the bright spot is relatively weak
compared to Hoard et al.'s \cite{HSSSB} Doppler maps and delayed in the disc
with respect to the expected position, where the ballistic stream hits
the accretion disc. The stream matter apparently enters the disc in
such a way that it first travels with the rotating material in the
disc before it releases and radiates its energy away. It coincides
possibly with Hoard et al.'s absorbing wall, where the line emission
diminishes because the material is nearly optically thick.

Baptista et al. \cite{BSC} estimate of the mass and radius of the
secondary fits to a main sequence star in the range M3.5 to
M4. Subtracting the flux of such a star from the uneclipsed component
requires the object to be later than about M3.7. Such a star would
leave us with no significant flux in I and a spectrum that rises
towards shorter wavelenghts and peaks in the B filter. It is not
possible to fit a black body spectrum to this distribution. The
uneclipsed flux must therefore originate in an extended, possibly
optically thick region with varying temperature. Contrary to HT~Cas
(Sect.~\ref{htcas}) and V2051~Oph (Sect.~\ref{v2051oph}), there is
no cool component of the uneclipsed flux.

Further details concerning this study including an analysis of all
of Baptista's individual light curves in high and low states will appear in
Vrielmann~\cite{V20U}. This presented study shows in particular, that
our PPEM analysis gives results in agreement with other methods,
especially concerning the distance to the system.

\begin{figure}
\begin{center}
\includegraphics[width=.48\textwidth]{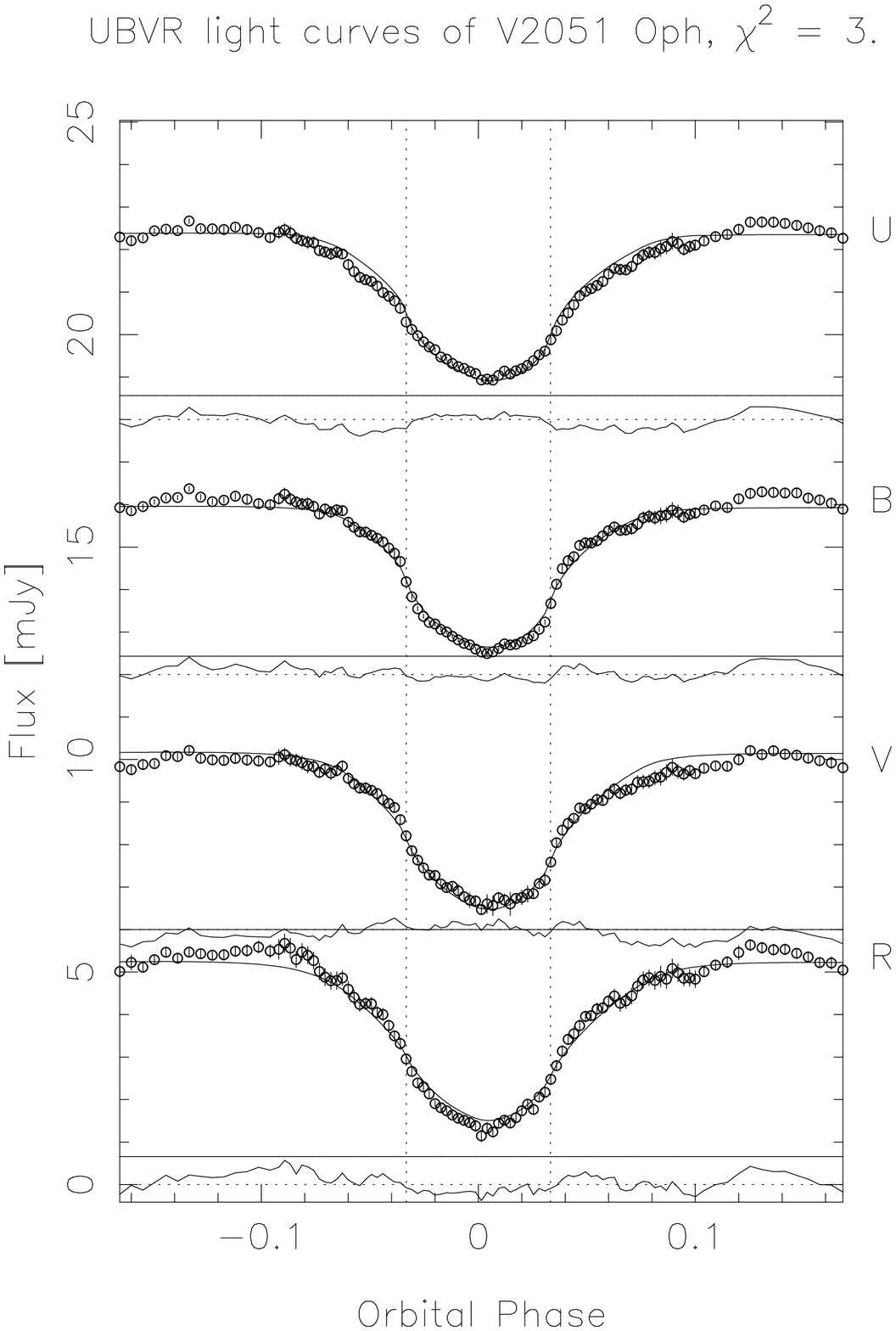}
\includegraphics[width=.48\textwidth]{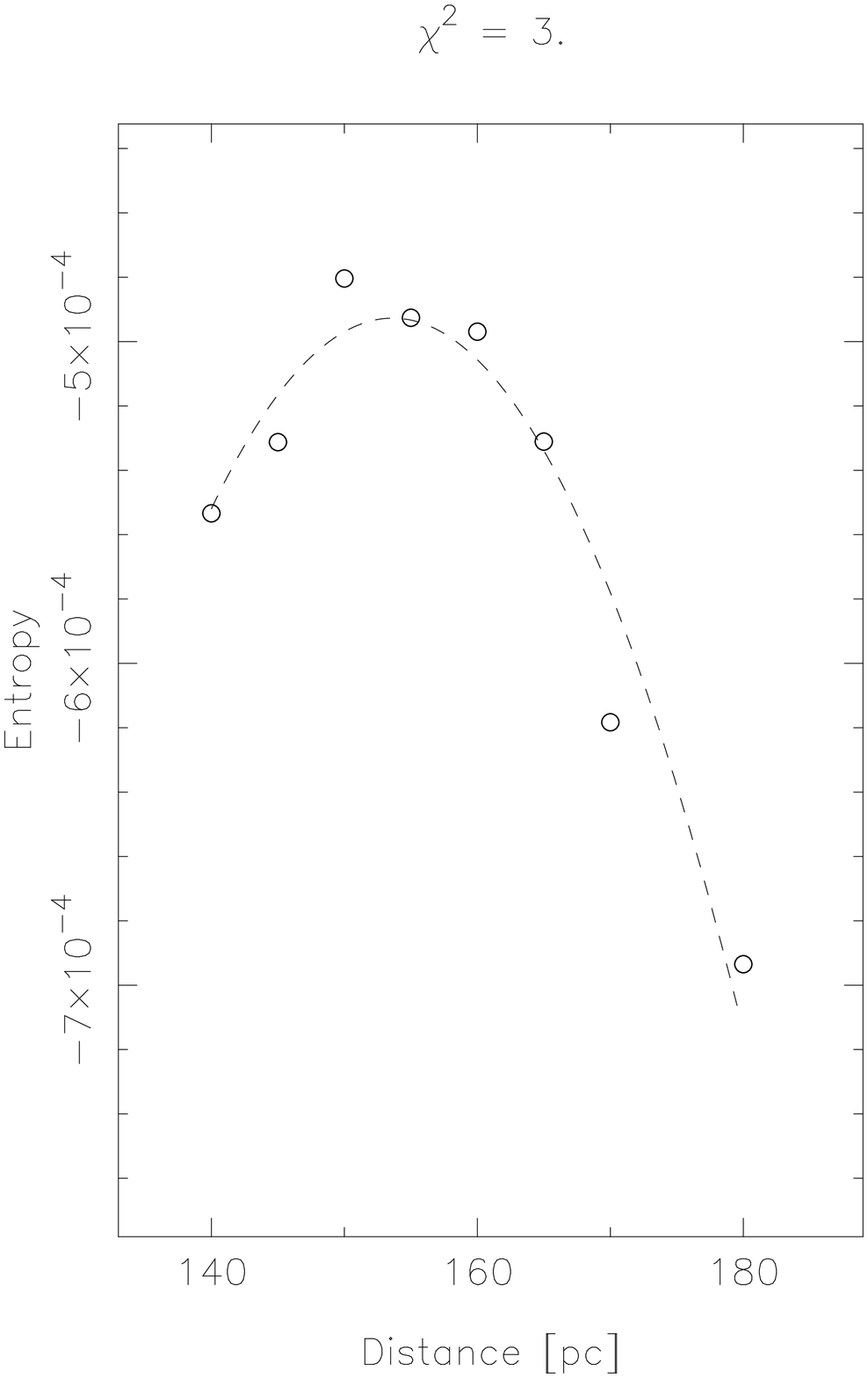}
\end{center}
\caption[]{{\em Left:} Averaged UBVRI light curves of {\bf V2051~Oph}
with the PPEM fits. For the explanation of the dashed and dotted lines
see Fig.~\ref{uu_f} (left). {\em Right:} Entropy vs.\ trial distance for
$\chi^2 = 3$ and a parabolic fit to the points peaking at 154~pc
\label{oph_f}}
\end{figure}

%%%%%%%%%%%%%%%%%%%%%%%%%%%%%%  V2051 Oph  %%%%%%%%%%%%%%%%%%%%%%%%%%%%%%%
\subsection{V2051~Ophiuchi}
\label{v2051oph}

The nature of the cataclysmic variable V2051~Oph was mysterious since
the discovery by Sanduleak \cite{S72}. It has been classified as a
(low-field) polar \cite{BW} where the orbital hump is
explained as a flaring accretion column \cite{WO}
and as a dwarf nova showing outbursts, double peaked emission lines
and an eclipse light curve that can be explained as caused by an
accretion disc \cite{BCHZ,WC,WBHGMC}. 
%(Warner \& Cropper 1983, Watts et al.\ 1986, Baptista et al.\ 1998) .

\begin{figure}
\begin{center}
\includegraphics[width=.48\textwidth]{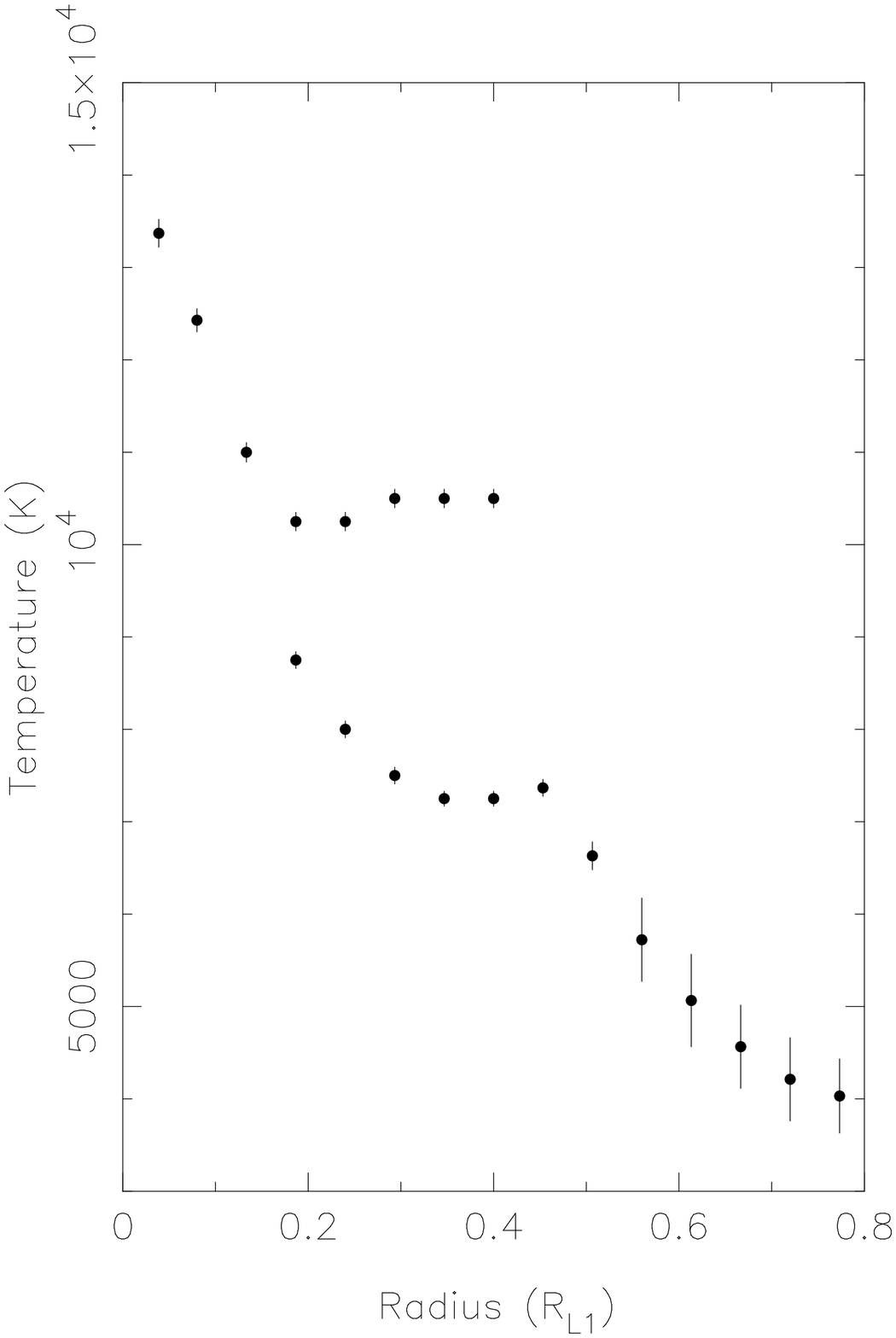}
\includegraphics[width=.48\textwidth]{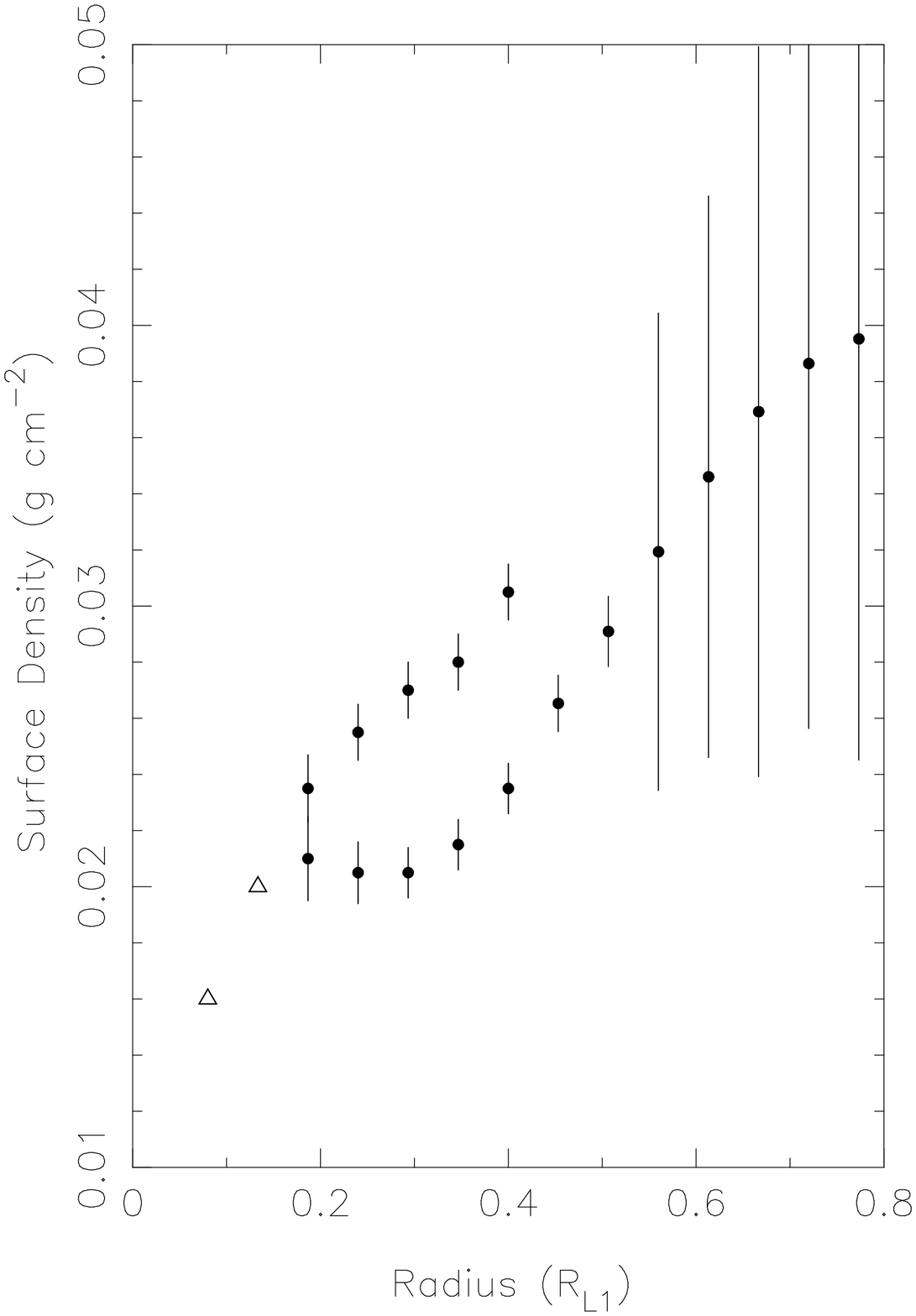}
\end{center}
\caption[]{{\bf V2051~Oph:} The averaged reconstructed temperature
({\em left}) and surface density ({\em right}) distributions with
error bars according to a study of the model.  The error bars
therefore indicate the reliability of the reconstructed parameter
values, not the variation in azimuth.  The values at intermediate
radii ($0.2\Rl$ to $0.4\Rl$) have been split up to illustrate the
range of parameter values.  At a radius of 0.04$\Rl$ the disc
is completely optically thick, i.e. we omitted its reconstructed
$\Sigma$ value. For the next two annulli the disc is nearly optically
thick, therefore we only give a lower limits for the surface density,
indicated by a triangles.
\label{oph_a}}
\end{figure}

It is certain that it is a high inclination system, displaying
eclipses and double peaked emission lines and after the super
outbursts in May 1998 and July 1999 it seems clear that the system
must have an accretion disc. Warner \cite{W96} suggested that it might be
the first system of a new class he terms polaroid. Polaroids are
similar to intermediate polars (IPs), i.e.\ dwarf novae in which the
white dwarf has an intermediately strong magnetic field that disrupts
the inner part of the disc. However, while IPs have a fast spinning
white dwarf, the primary in a polaroid is synchronized.

Figure~\ref{oph_f} (left) shows UBVRI light curves of V2051~Oph averaged
over eight eclipses and as used in the PPEM analysis of
Vrielmann~\cite{V20V}. Most flares and the flickering are averaged out,
except for the originally strong flare at phase 0.12.

We used the PPEM method to establish the first reliable distance
estimate to V2051~Oph as described in Sect.~\ref{distance}. For each
trial distance the data have been fitted to a $\chi^2$ of 3.
Figure~\ref{oph_f} (right) shows the resulting entropy as a function of trial
distance. A parabolic fit to the data points peaks at 154~pc.

\begin{figure}
\begin{center}
\includegraphics[width=.6\textwidth]{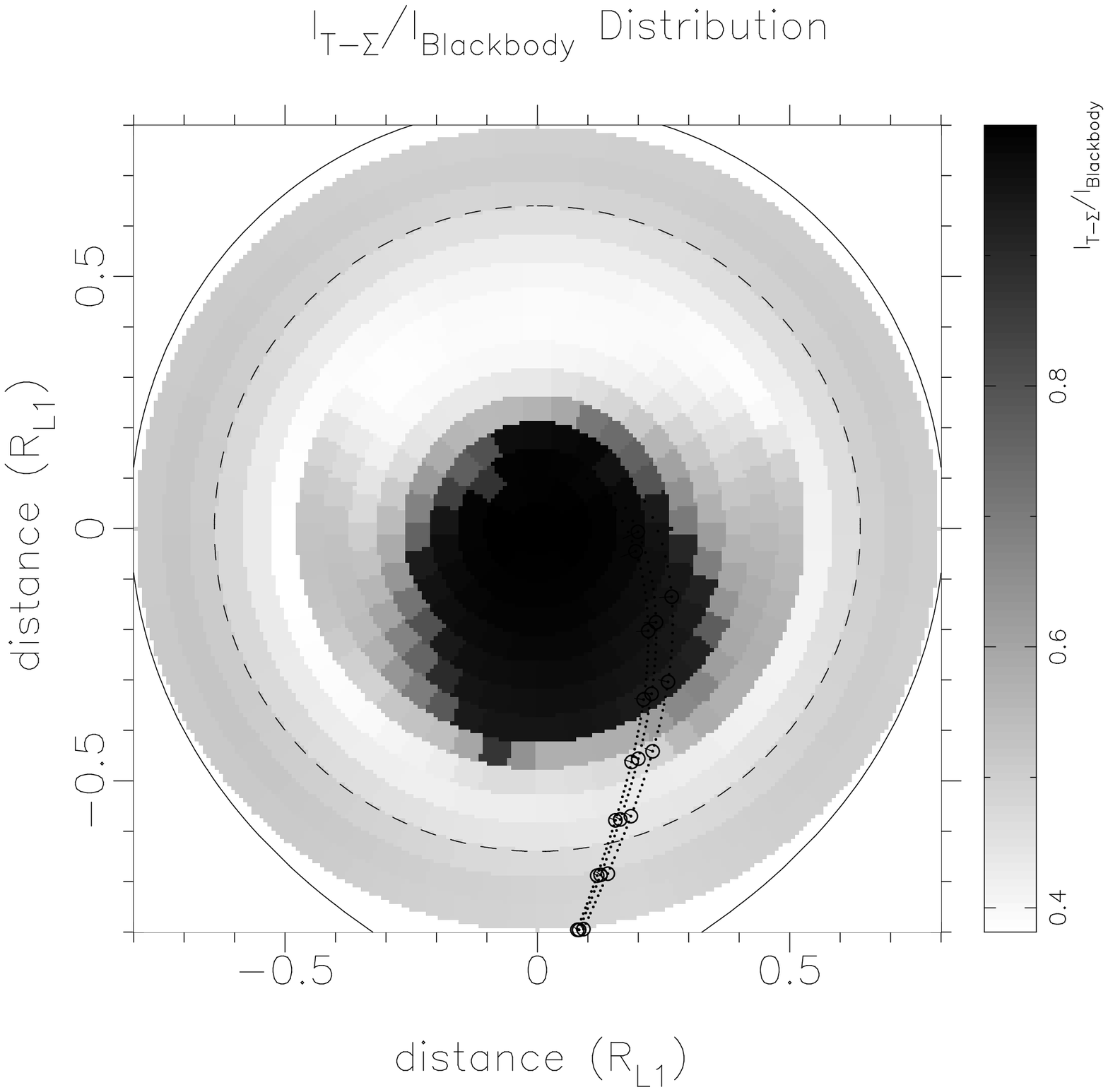}
\includegraphics[width=.38\textwidth]{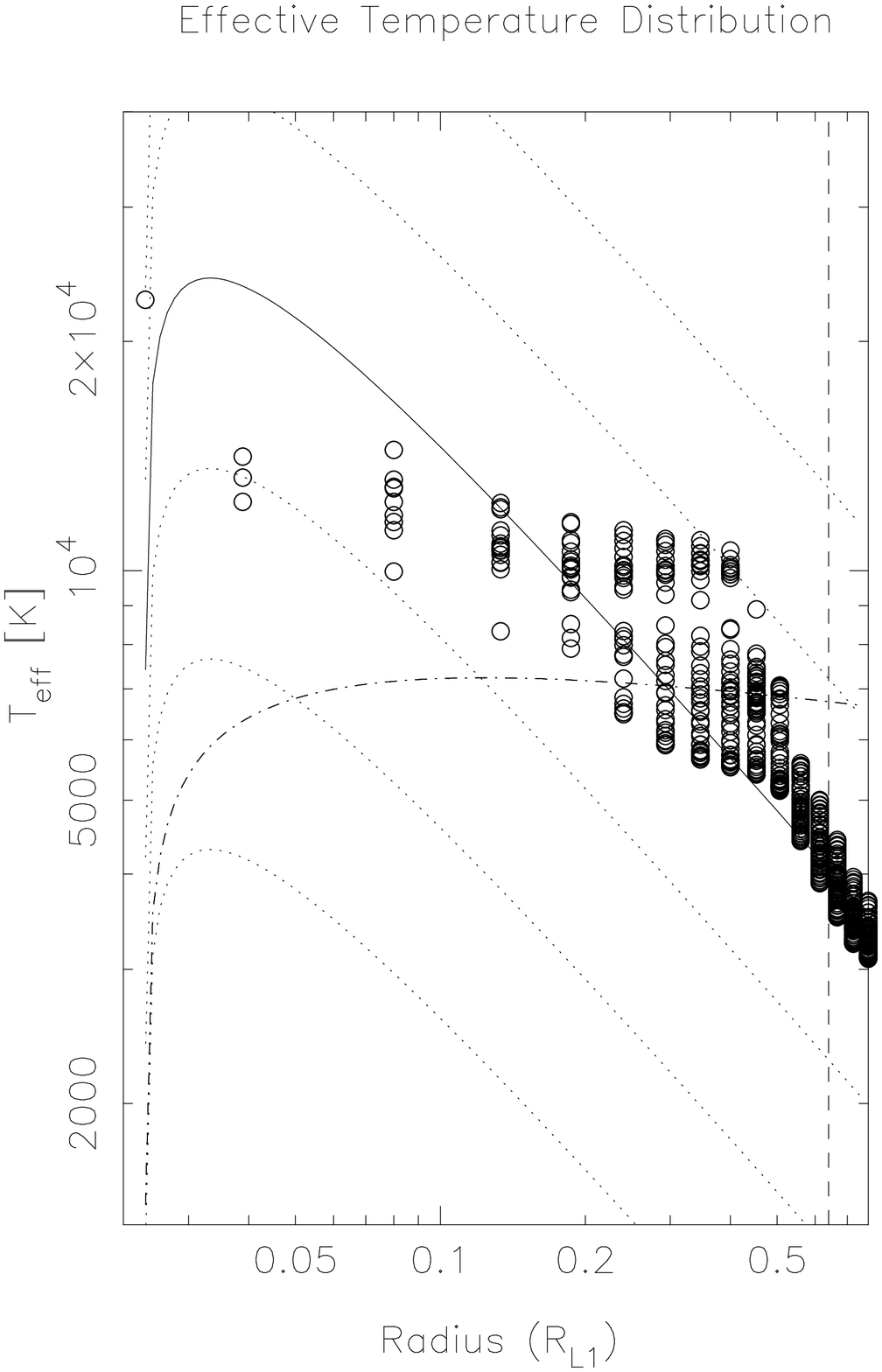}
\end{center}
\caption[]{{\bf V2051~Oph:} {\em Left:} The ratio
$I(T,\Sigma)/I_{BB}(T)$ as a gray-scale plot. {\em Right:} The radial
effective temperature distribution. The dashed and dotted lines in
both plots have the same meaning as in Fig.~\ref{uu_teff}
\label{oph_teff}}
\end{figure}

Figure~\ref{oph_a} shows averages of the reconstructed temperature and
surface density distribution with error bars according to a study of
the model. The values in the radial region $0.2\Rl$
to $0.4\Rl$ have been split up in an upper and lower branch because of
a true spatial separation of regions and to illustrate the range of
reconstructed parameters as well as that of the error bars. The upper
branch values in both parameters correspond to a relatively confined
region in the disc that we call {\em hot region}. It is located
between the stars and is nearly optically thick. The lower branch values
correspond to the remaining disc (cf.\ Fig.~\ref{oph_teff}, left).
In the outer regions of the disc the error bars in both parameters
increase, because the parameter values enter the region with {\em
banana} shaped $\chi^2$ contour lines (see Fig.~\ref{fig_xi2}).

The temperature $T$ decreases radially, as expected. In contrast, the
surface density $\Sigma$ rises towards the disc edge, but according to
the large error bars in the outer disc regions, $\Sigma$ could also be
almost constant throughout the disc. However, a decrease of $\Sigma$
with radius appears out of the question.

Using white dwarf spectra, the white dwarf temperature was
reconstructed to 22\,700~K. This value is close to typical values of
white dwarf temperatures, but somewhat higher than in the similar
objetcs HT~Cas, OY~Car and Z~Cha of around 15\,000~K \cite{GK}.
%Koester 1999 and references therein).

Figure~\ref{oph_teff} (left) gives a gray-scale display of the spatial
distribution of the ratio $I(T,\Sigma)/I_{BB}(T)$. The central parts up
to a radius of $0.15\Rl$ and the {\em hot region} between the stars
are optically thick. In the outer region, $I(T,\Sigma)/I_{BB}(T)$ is
close to unity, because both $I(T,\Sigma)$ and $I_{BB}(T)$ vanish there.
Part of the optically thick {\em hot region} can be attributed to the
bright spot, expected at the point where the ballistic gas stream hits
the disc. But the {\em hot region} is too large and spreads out too far
towards negative azimuths (into the following lune).

The right panel of Fig.~\ref{oph_teff} shows the radial distribution of the
effective temperature in V2051~Oph's disc. A clear separation can be
seen for the {\em hot region} and the remainder of the disc. While the
mass accretion rate $\Md$ of the {\em hot region} lies above the critical
value, the remaining disc's $\Md$ is approximately equal to the
critical value or below. Since V2051~Oph is a dwarf nova showing
occational outbursts, this mass accretion rate is far too high. It
should lie well below the critial value.

A grayscale plot of the $\Teff$ distribution (not shown here) shows a
distinct ridge within the {\em hot region}. It is almost parallel to
the binary axis, but tilted by about 15$^\circ$ towards the following
lune of the disc.

In order to explain the {\em hot region}, the presence of an
outbursting disc and the variable humps as seen by Warner \&
O'Donoghue \cite{WO} we propose the model for V2051~Oph as described in
Sect.~\ref{polaroid}.

Calculating the viscosity parameter $\alpha$ we derive values far too
large. While theory predicts values in the range of $\alpha \sim
0.01$ (see Sect.~\ref{viscosity}), our values lie between 30 and
1000. In Sect.~\ref{discmodel} we explain how we can solve this
discrepancy with our proposed model of accretion discs.

The uneclipsed component is reconstructed to 0.56~mJy, 0.43~mJy,
0.00~mJy and 0.66~mJy for the filters UBVR, respectively.  Baptista et
al.'s~\cite{BCHZ} mass and radius for the secondary, fits to a M4.5 main
sequence star \cite{KM}, however, this leads to
fluxes of 0.07~mJy in V and 0.22~mJy in R. The secondary must
therefore be of slightly later type or the error in the reconstructed
values is of order a few hundredth mJy.

The separation of the UV and IR component means the uneclipsed
component must originate in two separate sources. The UV flux probably
comes from the hot chromosphere, extending to at least a few white
dwarf radii above the disc plane. The IR source may be a cool disc
wind. These estimates, though, are only valid as long as the spectral
model is a good representation of the true emissivity of the disc.

Any further details about this analysis are described in Vrielmann~\cite{V20V}.

%%%%%%%%%%%%%%%%%%%%%%%%%%%%%%%  HT Cas  %%%%%%%%%%%%%%%%%%%%%%%%%%%%%%%
\subsection{HT~Cassiopeiae}
\label{htcas}

HT~Cas is an unusual dwarf nova in that it does not show much of a
bright spot and has occationally very long quiescent times of up to 9
years. On the other hand, Patterson \cite{P} suggested it might serve as
a {\em Rosetta stone} in explaining what drives the accretion
discs. Since it is one of the few eclipsing dwarf novae it gives us a
special opportunity for the analysis of a quiescent disc with PPEM.

\begin{figure}
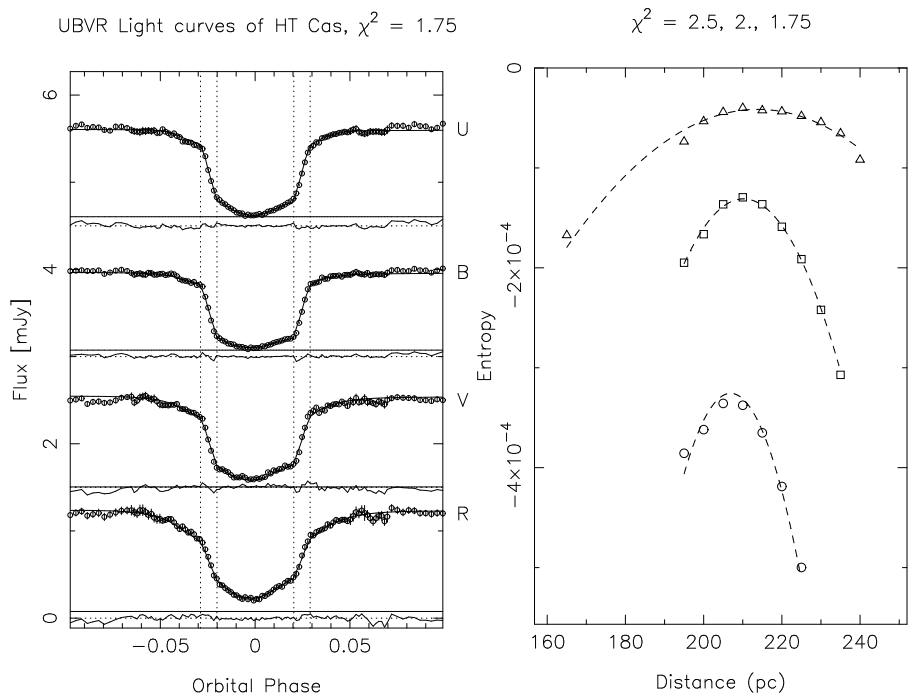

\begin{center}
\includegraphics[width=.5\textwidth]{ht205_175f.epsi}
\includegraphics[width=.47\textwidth]{ht205_175en.epsi}
\end{center}
\caption[]{{\em Left:} Averaged UBVRI light curves of {\bf HT~Cas}
with the PPEM fits. For the explanation of the dashed and dotted lines
see Fig.~\ref{uu_f} (left). {\em Right:} Entropy vs.\ trial distance
for $\chi^2 = 1.75$ and a parabolic fit to the points peaking at
207~pc
\label{ht_f}}
\end{figure}

We used UBVR light curves previously analysed and published by Horne
et al.~\cite{HWS}. Figure~\ref{ht_f} (left) shows the averaged UBVR eclipse
light curves as prepared for our PPEM analysis \cite{VHH2}.
Most of the flickering has disappeared and the most
prominent remaining feature is the white dwarf eclipse.

We used these data to determine an independent distance estimate as
described in Sect.~\ref{distance}. Previous reliable estimates
involved the secondary \cite{M} and the white dwarf \cite{WNHR}
and resulted in values of 140~pc and 165~pc, respectively.
Our PPEM estimate is significantly larger with a value of 207~pc
(Fig.~\ref{ht_f}, right), however, the fits to the light curves are
very good with a $\chi^2$ of 1.75. Sect.~\ref{patchy} describes our
suggested solution for this distance problem. It allows us to use the
parameter values as reconstructed.

\begin{figure}
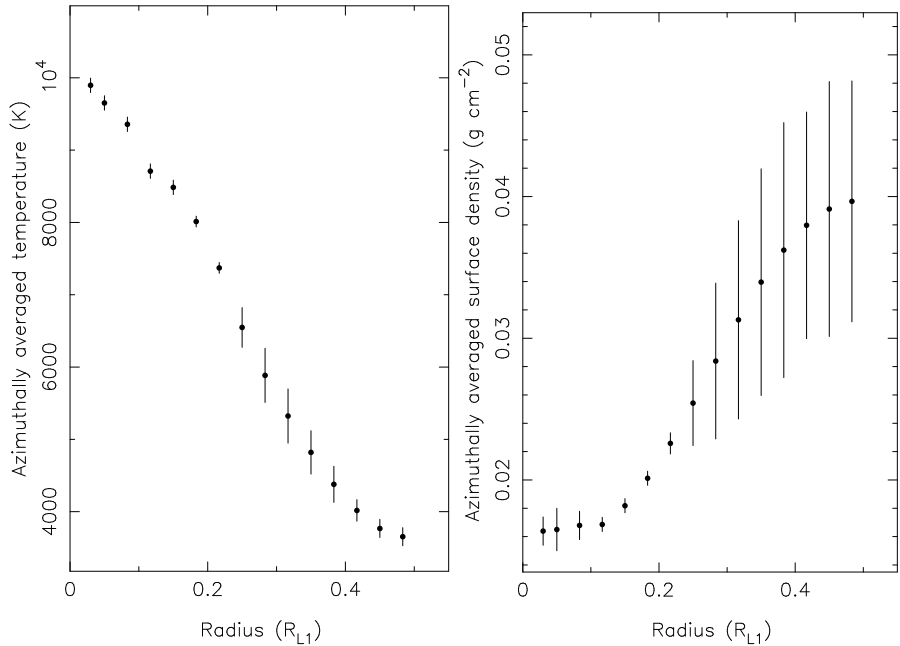

\begin{center}
\includegraphics[width=.48\textwidth]{ht205_175ae1.epsi}
\includegraphics[width=.48\textwidth]{ht205_175ae2.epsi}
\end{center}
\caption[]{{\bf HT~Cas:} The averaged reconstructed temperature
({\em left}) and surface density ({\em right}) distributions with
error bars according to a study of the model
\label{ht_a}}
\end{figure}

Figure~\ref{ht_a} gives azimuthally averaged temperature and surface
density distributions as reconstructed for the trial distance
205~pc. Like in the disc of V2051~Oph, the temperature decreases with
radius. However, the surface density clearly increases radially even
within the large error bars. The corresponding intensity distribution
allow us to set a radius of the disc at $0.4\Rl$.

The reconstructed temperature of the white dwarf is 22\,600~K. Because
of our larger distance, this value is larger than Wood et al.'s~\cite{WNHR}
value of $18\,700\pm1\,800$~K determined from the same data.

\begin{figure}
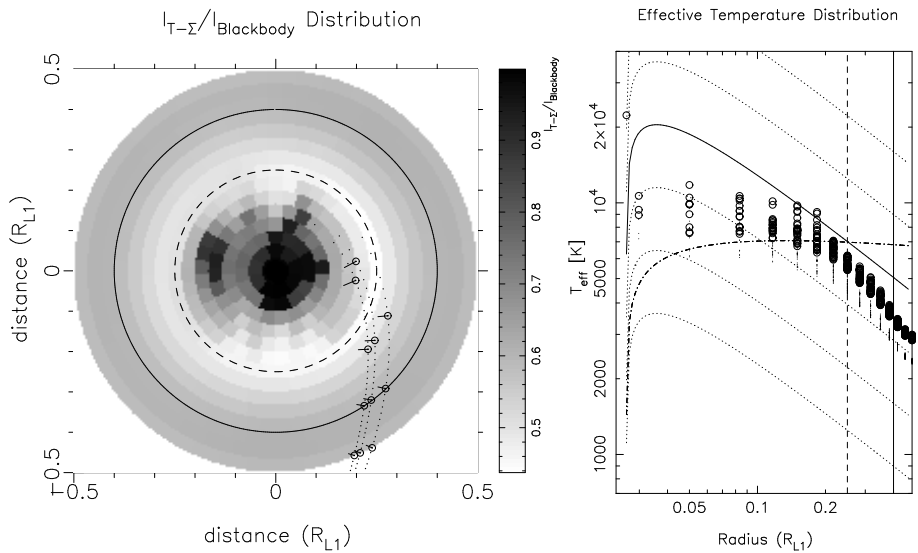

\begin{center}
\includegraphics[width=.6\textwidth]{ht205_175iv.epsi}
\includegraphics[width=.38\textwidth]{ht205_175tn.epsi}
\end{center}
\caption[]{{\bf HT~Cas:} {\em Left:} The ratio $I(T,\Sigma)/I_{BB}(T)$
as a gray-scale plot. The solid circle is the disc radius, the dashed
circle indicates where the surface density values become uncertain.
The Roche-lobe of the primary lies entirely outside the displayed
area. Otherwise same as Fig.~\ref{uu_teff} (left).
{\em Right:} The radial effective temperature distribution for a
distance of 205~pc (circles) and 133~pc (dots). The dashed
vertical line gives the radius where the surface density values become
uncertain and the solid line gives the disc radius. Otherwise same as
Fig.~\ref{uu_teff} (right)
\label{ht_teff}}
\end{figure}

As shown by Fig.~\ref{ht_teff} (left), only the very central parts of
the accretion disc are optically thick and the $I(T,\Sigma)/I_{BB}(T)$
ratio shows some asymmetry, also present in the original $T-\Sigma$
maps.  We would expect the emission lines to originate only in regions
of the disc with radii $r > 0.05\Rl$.

Assuming a distance of 205~pc, the effective temperature is
significantly larger than the critical values (see Fig.~\ref{ht_teff},
right). It would be difficult to explain such high effective
temperature and therefore mass accretion rates in a system that shows
only rare outbursts, i.e.\ should have an exceptionally low mass
accretion rate. If we agree with the cited literature values for the
distance and assume a distance of 133~pc (see Sect.~\ref{patchy}),
the effective temperature would drop to values below or close to the
critical temperatures also shown in Fig.~\ref{ht_teff} (right). In the
inner part of the disc ($r < 0.2\Rl$), the $\Teff$ is flat, while
outside this range the disc is steady state like with a mass accretion
rate of about $2\times10^{15}$\gs.

Similarly as in the case of V2051~Oph, the viscosity parameter
$\alpha$ we determine from the maps using Eq.~(\ref{eq_viscosity})
assumes values too large by three to four orders of magnitude.
Sect.~\ref{discmodel} describes our model for the disc in both
systems which qualitatively explains our solution of the discrepancy
to theory.

The reconstructed uneclipsed component can partly be explained by a
M5.4 secondary \cite{M}. Subtracting such a main sequence star
using a distance of 133~pc, basically no extra uneclipsed flux in the
V band remains, like in V2051~Oph. However, in U, B, and R we find
significant additional flux of 0.1~mJy, 0.08~mJy, and 0.05~mJy,
respectively. As in V2051~Oph this must originate from two distinctly
different sources, presumeably a hot chromosphere (providing the U and
B flux) and a cool disc wind, giving rise to the R flux.

%%%%%%%%%%%%%%%%%%%%%%%%%%%%%  Discussion  %%%%%%%%%%%%%%%%%%%%%%%%%%%%%

\section{Discussion}
\label{discussion}
\subsection{About the nature of V2051~Oph}
\label{polaroid}
V2051~Oph has often been compared to HT~Cas, OY~Car and Z~Cha
because of the similar orbital periods and all being SU~UMa
stars. Especially the similarly long and similarly variable outburst
intervals of HT~Cas and V2051~Oph is striking. One would hope that if
one could explain one of these objects, all were understood.

However, V2051~Oph has some special features, like the variable hump
as seen in the light curves of Warner \& Cropper \cite{WC} and Warner \&
O'Donoghue \cite{WO} that can only be explained by a hump source very
close to the white dwarf. Objects with an (disrupted) accretion disc
and an (flaring) accretion column would most easily be explained as an
intermediate polar (IP). However, IPs usually show characteristic
X-ray emission. V2051~Oph has only been detected as a very faint X-ray
source, but on the other hand no high inclination system shows high
X-ray count rates. Instead there seems to exist an anti-correlation
between the observable X-ray emission measure and the inclination
angle. The X-ray source therefore is very close to the white dwarf and
must be obscured or absorbed in the high inclination systems 
\cite{HCP,vBV}.
%(Holcomb et al.\ 1994, van Teeseling et al.\ 1996).

The absence of circular polarisation in the wavelength range
3500-9200~\AA \cite{C} could 
be explained by a very low magnetic field of 1~MG or less~\cite{WO}, as
expected for IPs.

If V2051~Oph is indeed an intermediate polar, the inner disc radius
must be very small, of order less than $0.1~\Rl$. Otherwise, it would
have been reconstructed by PPEM. On the other hand, the accretion
curtains would partially fill out the space between the white
dwarf and the accretion disc. 

Warner \cite{W96} suggests that V2051~Oph could be one of the systems he
terms polaroid. A polaroid is an intermediate polar with a
synchronized white dwarf. With our analysis we cannot distinguish
between an IP and a polaroid, but oscillation found in V2051~Oph
indicate a non-synchronous white dwarf \cite{S20}
and therefore we favour the IP model.

It remains to be explained what causes the {\em hot region} in the
disc. One possibility is limb brightening of the accretion disc that
we did not take into account. If the disc was flared this effect would
even be enhanced. However, it would not explain the slightly
asymmetric ridge we see in the effective temperature
distribution. Another possibility is an illumination from a bulge
where the ballistic stream hits the accretion disc as described in
Buckley \& Tuohy \cite{BT}. However, this would also make it difficult to
explain the ridge in the effective temperature map.  A third
explanation is a warp in the disc in such a way that the disc region
between the stars is illuminated by the white dwarf at azimuths
$-15^\circ$ and the region on the opposite side of the disc ($+
165^\circ$) has a bulge so that the outer parts of the disc cannot be
illuminated by the central object (for a more detailed explanation see
\cite{V20V}). However, it would be difficult to explain if
it is stable in the rotating frame of the binary.

A PPEM analysis of multi-colour light curves from other time periods
would be helpful in deciding what causes the {\em hot region}, if its
location in the disc is stable, rotates in the disc or is variable in
its presence.

\subsection{Accretion Disc model for HT~Cas and V2051~Oph}
\label{discmodel}
Instead of dismissing Eq.~(\ref{eq_viscosity}) or our PPEM analysis
altogether, we were seeking for an explanation for the high values of
$\alpha$ of a few hundreds that we derive for the disc in both HT~Cas and
V2051~Oph. 

If we compare our derived surface density values with the values used
in the disc instability model \cite{FLP,LMR,S82} 
%(e.g.\ Ludwig et al.\ 1994, Faulkner et al.\ 1983, Smak 1982) 
we find that our values are much lower than
those necessary for the disc entering the hysteresis curve. Theory
predicts values between
10 and 100 \gcm\ for $\alpha$'s between 0.1 and 1 and even at larger
$\Sigma$'s for smaller $\alpha$'s. Since the disc cannot change the
$\Sigma$ from quiescence to outburst that dramatically, this means the
outburst can only occur if the disc is very well optically thick
already before its onset. On the other hand line emission is seen
during rise and decline of an outburst \cite{HRNZ}.

To solve this problem we suggest that the disc consists of a cool,
optically thick layer carrying most of the matter and which is
sandwiched by a hot, optically thin chromosphere. The surface
densities in the underlying disc must be a factor $10^4$ ($\Sigma$ a
few hundreds) larger then the reconstructed values to bring the
$\alpha$'s down to reasonable values of between 0.01 and 0.1.
However, we only ``see'' the chromosphere, therefore we derive the
large values for $\alpha$.

The presence of a hot chromosphere is supported by the excess of blue
emission in the uneclipsed component in HT~Cas, V2051~Oph and UU~Aqr.
This emission must come from regions above the disc plane that are
never eclipsed, i.e.\ a few white dwarf radii. This means, the
chromosphere is an extended layer on top of the disc
surface. Furthermore, this chromosphere must be the location where the
line emission originates.

\subsection{HT~Cas' patchy disc}
\label{patchy}
In a similar way as above, the distance problem with HT~Cas does not
mean we have to dismiss the PPEM algorithm right away. As we have
shown, the PPEM estimate gives results in agreement with literature
values for UU~Aqr (see Sect.~\ref{uuaqr}). Furthermore, our use of
literature values for IP~Peg and VZ~Scl does not indicate any
disagreement.

Instead, we can explain our large discrepancy between our PPEM
distance estimate and literature values for HT~Cas by allowing the
disc to be {\em patchy}, i.e.\ the emitting surface is smaller then
the geometrical surface. Using Marsh \cite{M} distance estimate with a
recalibrated Barnes-Evans relation (for details see \cite{VHH2}),
the covering factor must be of the order $C = (133/207)^2=41$\%.

The patchy disc can be caused by magnetic activity in localized
regions on the surface or in the upper layers of the accretion disc.
Magnetic flux created by dynamo action and/or the Balbous-Hawley
instabilities driving the viscosity rises out of the cool midplane
regions and dissipates most of the viscously generated energy via
magnetic reconnection or similar coronal processes.  This model also
explains why the energy dissipation rate is proportional to the local
orbital frequency \cite{HS}.

Our PPEM analysis of HT~Cas gives an indirect evidence of the
hydromagnetic nature of the anomalous viscosity in accretion disks.

\section{Summary}

We have shown that Physical Parameter Eclipse Mapping is a powerful
tool that helps us to learn about the physics of accretion discs in
cataclysmic variables. This review can only highlight some interesting
results. But, whether we derive ``just'' an independent distance estimate or
get insight into the disc structure, there is always something to learn
from the application of this method.

\section*{Acknowledgments}

This work was funded by the South African NRF and CHL Foundation.
I wish to thank Raymundo Baptista and Keith Horne for communicating me
their data to apply my PPEM method. Many thanks go as well to Keith,
Rick Hessman, Stephen Potter and Brian Warner for fruitful
discussions.  Futhermore, I am very grateful to Hilde Langenaken for
babysitting my daughter Nina in Brussels and thank Nina for
giving me so much joy and being so cooperative during the workshop.

%%%%%%%%%%%%%%%%%%%%%%%%%%%%%  References  %%%%%%%%%%%%%%%%%%%%%%%%%%%%%

%INDEX%%%%%%%%%%%%%%%%%%%%%%%%%%%%%%%%%%%%%%%%%%%%%%%%%%%%%%%%%%%%%%%
% Please check with the editor of your book whether he plans to
% include a "mutual" subject index - if so, please code your entries
% in the standard syntax. For your own purposes you may print your
% "personal" index by using the following commands:
%
%\clearpage
%\addcontentsline{toc}{section}{Index}
%\flushbottom
%\printindex
%%%%%%%%%%%%%%%%%%%%%%%%%%%%%%%%%%%%%%%%%%%%%%%%%%%%%%%%%%%%%%%%%%%%%

\end{document}